\documentclass[a4paper,aps,11pt,nofootinbib,pdftex]{revtex4}
\usepackage{amsmath}
\usepackage{color}
\usepackage{ulem}
\usepackage{ascmac}
\usepackage{setspace}

\usepackage{graphicx}

\begin{document}

\title{Charge Screened Non-Topological Solitons \\
in a Spontaneously Broken U(1) Gauge Theory}

\begin{spacing}{1.0}
\begin{flushright}
\hfill{OCU-PHYS 491}
\\
\hfill{AP-GR 151}
\\
\hfill{NITEP 3}
\end{flushright}
\end{spacing}


\author{Hideki Ishihara}
\email{ishihara@sci.osaka-cu.ac.jp}
\author{Tatsuya Ogawa}
\email{taogawa@sci.osaka-cu.ac.jp}
\affiliation{
 Department of Mathematics and Physics,
 Graduate School of Science,
Nambu Yoichiro Institute of Theoretical and Experimental Physics (NITEP),
Osaka City University, Osaka 558-8585, Japan}

\begin{abstract}
We construct, numerically, stationary and spherically symmetric nontopological soliton
solutions in the system composed of a complex scalar field, a U(1) gauge field,
and a complex Higgs scalar field that causes spontaneous symmetry braking.
It is shown that the charge of the soliton is screened by counter charge everywhere.
\end{abstract}

\maketitle

\section{Introduction}
Nontopological solitons, which are energy minimum solutions
under the condition of fixed conserved U(1) charge in classical field theories,
appear in various theories:
a coupled system of a complex scalar field and a real scalar field \cite{Friedberg_Lee_Sirlin},
a complex scalar field with a nonlinear self interactions \cite{Coleman},
and so on\footnote{
Potentials inspired by the super symmetric theories also allowed the nontopological
soliton solutions \cite{Kusenko:1997zq, Kasuya:2000sc}.}.
In the system with the gauge symmetry, the nontopological soliton solutions
are also studied \cite{Lee_Stein-Schabes_Watkins_Widrow, Shi_Li, Gulamov_etal}.

The gauge theory with spontaneous symmetry breaking is the most fundamental
framework in the modern physics. We present, in this article, nontopological
soliton solutions in the system composed of a complex scalar field coupled to
a U(1) gauge field, and a complex Higgs scalar field that causes the spontaneous symmetry breaking.
This is a generalization of Friedberg-Lee-Sirlin's model \cite{Friedberg_Lee_Sirlin, Shi_Li}.

We show that the charge of the nontopological solitons
are perfectly screened \cite{Ishihara:2018eah}, namely,
the charge density carried by the complex scalar field
is canceled out by the counter charge cloud carried by the other fields everywhere.
In contrast to the known fact
that the mass of the gauged nontopological soliton whose charge is not screened
is bounded above\cite{Lee_Stein-Schabes_Watkins_Widrow, Shi_Li, Gulamov_etal},
we show the charge screened nontopological solitons can have any large amount of mass.

\section{Basic System}

We consider the system described by the action
\begin{align}
	S=\int d^4 x \left(
		-(D_{\mu}\psi)^*(D^{\mu}\psi) -(D_{\mu}\phi)^*(D^{\mu}\phi)
		-V(\phi)-\mu \psi^{\ast} \psi \phi^{\ast}\phi -\frac{1}{4}F_{\mu\nu}F^{\mu\nu}
	\right),
\label{eq:action}
\end{align}
where $\psi$ is a complex scalar field,
$\phi$ is a complex Higgs scalar field with the potential
\begin{align}
  	V(\phi)=\frac{\lambda}{4}(\phi^{\ast}\phi-\eta^2)^2,
 \label{eq:potential}
\end{align}
where $\lambda$ and $\eta$ are positive constants,
and $F_{\mu\nu}:=\partial_{\mu}A_{\nu}-\partial_{\nu}A_{\mu}$ is the field strength of
a U(1) gauge field $A_{\mu}$.
Here, $D_{\mu}$ in \eqref{eq:action} is the covariant derivative defined by
\begin{align}
  	D_{\mu}\psi :=\partial_{\mu}\psi -ieA_{\mu}\psi, \quad
	D_{\mu}\phi :=\partial_{\mu}\phi -ieA_{\mu}\phi,
 \label{eq:covariant_derivative}
\end{align}
where $e$ is a coupling constant.

The action \eqref{eq:action} is invariant under the local U(1) transformation
and the global U(1) transformation:
\begin{equation}
\begin{split}
  	&\psi(x) \to \psi'(x)=e^{i(\chi(x)-\gamma)}\psi(x),
\\
  	&\phi(x) \to \phi'(x)=e^{i(\chi(x)+\gamma)}\phi(x),
\\
 	&A_{\mu}(x)\to A_{\mu}'(x)=A_{\mu}(x)+e^{-1}\partial_{\mu}\chi(x),
\end{split}
 \label{eq:gauge_tr}
\end{equation}
where $\chi(x)$ is an arbitrary function and $\gamma$ is a constant.
Owing to the invariance, the system has conserved currents
\begin{align}
  j_\psi^{\nu} &:=i e \left(\psi^{\ast}(D^{\nu}\psi)-\psi(D^{\nu}\psi)^{\ast}\right),
  \label{eq:j_psi}
\\
  j_\phi^{\nu} &:=i e \left(\phi^{\ast}(D^{\nu}\phi)-\phi(D^{\nu}\phi)^{\ast}\right),
  \label{eq:j_phi}
\end{align}
hence, total charge of $\psi$ and $\phi$ defied by
\begin{align}
  Q_\psi := \int \rho_\psi~ d^3x,
\quad
  Q_\phi := \int \rho_\phi~ d^3x
  \label{eq:Q}
\end{align}
are conserved,
where $\rho_\psi:=j_{\psi}^t $ and $\rho_\phi:=j_{\phi}^t $.

The energy of the system is given by
\begin{align}
 	E=\int d^3x \biggl( \left|D_{t}\psi\right|^2 & +(D_{i}\psi)^{\ast}(D^{i}\psi)
		 +\left|D_{t}\phi\right|^2+(D_{i}\phi)^{\ast}(D^{i}\phi)\notag \\
	  	& +V(\phi)+\mu|\psi|^2|\phi|^2 +\frac{1}{2}\left(E_iE^i+B_iB^i\right)\biggl) ,
\label{eq:energy}
\end{align}
where $E_i:=F_{i0}$,  $B^i:=1/2\epsilon^{ijk}F_{jk}$, and $i$ denotes spatial index.
In the vacuum state, which minimizes the energy \eqref{eq:energy},
$\psi$, $\phi$ and $A_{\mu}$ should satisfy
\begin{align}
 \psi=0,~ \phi^{\ast}\phi=\eta^2,~ \text{and}  ~  D_{\mu}\phi=0.
 \label{eq:VEV2}
\end{align}
Equivalently, the fields should take the form
\begin{align}
  \psi=0,~ \phi = \eta e^{i\theta(x)},~   \text{and}~  A_{\mu}=e^{-1}\partial_{\mu}\theta,
\label{eq:VEV}
\end{align}
where $\theta$ is an arbitrary function. Since the gauge field is pure gauge, then $F_{\mu\nu}=0$.
By the vacuum expectation value of the Higgs scalar field $\eta$,
the gauge field $A_\mu$ and the complex scalar field $\psi$ acquire the mass
$m_A=\sqrt{2}e \eta$ and $m_{\psi}=\sqrt{\mu}\eta$, respectively.
The real scalar field that denotes a fluctuation of the amplitude of $\phi$
around $\eta$ also acquires the mass $m_\phi=\sqrt{\lambda}\eta$.
In the vacuum state \eqref{eq:VEV}, a global U(1) symmetry still exists.

By varying (\ref{eq:action}) with respect to $\psi^{\ast}$, $\phi^{\ast}$, and $A_{\mu}$,
we obtain equations of motion:
\begin{equation}
\begin{split}
  &D_{\mu}D^{\mu}\psi-\mu \phi^{\ast}\phi \psi =0,
\\
  &D_{\mu}D^{\mu}\phi-\frac{\lambda}{2}\phi(\phi^{\ast}\phi-\eta^2)
		-\mu \phi \psi^{\ast}\psi =0,
\\
  &\partial_{\mu}F^{\mu\nu}- j_\phi^{\nu}- j_{\psi}^{\nu}=0.
\end{split}
\label{eq:basic_eq}
\end{equation}

We assume that the fields are stationary and spherically symmetric
in the form
\begin{equation}
\begin{split}
  	\psi=e^{i\omega t}u(r),
\quad
  	\phi=e^{i\omega' t}f(r),
\quad
 	&A_t=A_t(r),\quad \mbox{and}\quad A_i=0,
\end{split}
\label{eq:ansatz}
\end{equation}
where $\omega$ and $\omega'$ are constants, $u(r)$ and  $f(r)$ are real functions of $r$.
Using the gauge transformation \eqref{eq:gauge_tr} 
we fix the variables as
\begin{equation}
\begin{split}
  	&\phi(r) \to f(r),
\\
  	&\psi(t,r) \to e^{i\Omega t}u(r):=e^{i(\omega-\omega') t}u(r),
\\
 	&A_t(r) \to \alpha(r):= A_t(r)-e^{-1}\omega' ,
\end{split}
\label{eq:gauge_fixed}
\end{equation}
Substituting \eqref{eq:gauge_fixed}
into the field equations \eqref{eq:basic_eq}
we obtain
\begin{equation}
\begin{split}
 &\frac{d^2u}{dr^2}+\frac{2}{r}\frac{du}{dr}+(e\alpha-\Omega)^2u-\mu f^2u=0,
\\
 &\frac{d^2f}{dr^2}+\frac{2}{r}\frac{df}{dr}
		-\frac{\lambda}{2}f(f^2-\eta^2)+e^2 \alpha^2 f -\mu u^2 f=0,
\\
 &\frac{d^2\alpha}{dr^2}+\frac{2}{r}\frac{d\alpha}{dr}+\rho_\psi +\rho_\phi =0.
\end{split}
\label{eq:eq_of_motion}
\end{equation}
where the charge densities $\rho_\psi$, $\rho_\phi$ are given by
\begin{align}
	\rho_\psi = -2e(e\alpha-\Omega) u^2 ,
\quad
	\rho_\phi  = -2e^2\alpha f^2.
\label{eq:rho}
\end{align}
In the equations \eqref{eq:eq_of_motion},
the parameter $\Omega$ characterizes solutions.
We seek configurations of the fields with a non-vanishing value of $\Omega$.

The set of equations \eqref{eq:eq_of_motion} are derived
from the effective action in the form
\begin{align}
	&S_\text{eff} = \int  r^2 dr \left(
		 \biggl(\frac{du}{dr}\biggr)^2 + \biggl(\frac{df}{dr}\biggr)^2 -\frac12 \biggl(\frac{d\alpha}{dr}\biggr)^2
		- U_\text{eff}	\right),
\label{eq:S_eff}
\\
	&U_\text{eff}:= -\frac{\lambda}{4}(f^2-\eta^2)^2-\mu f^2 u^2
		+(e\alpha-\Omega)^2u^2 +e^2f^2 \alpha^2.
\end{align}
If we regard that the coordinate $r$ resembles \lq time\rq, the effective
action \eqref{eq:S_eff} describes a mechanical system of three degrees of freedom,
$u$, $f$ and $\alpha$, where the \lq kinetic\rq\ term of $\alpha$ has the wrong sign.

Using the ansatz \eqref{eq:gauge_fixed},
we rewrite the energy \eqref{eq:energy} for the symmetric system as
\begin{align}
	E=4\pi \int_0^{\infty}r^2 dr
		\left(	\biggl(\frac{du}{dr}\biggr)^2 \right. &+ \biggl(\frac{df}{dr}\biggr)^2
	+\frac{1}{2}\biggl(\frac{d\alpha}{dr}\biggr)^2
\cr
	& \left. +\frac{\lambda}{4}(f^2-\eta)^2+\mu f^2u^2 +(e\alpha-\Omega)^2u^2 +e^2f^2\alpha^2
	  \right).
\label{eq:energy2}
\end{align}
At the origin, we impose the regularity conditions for the spherically symmetric fields as
\begin{align}
  \frac{du}{dr}\to 0 , ~ \frac{df}{dr}\to 0  , ~
  \frac{d\alpha}{dr}\to 0 \quad \mbox{as}\quad r\to 0.
\label{eq:BC_origin}
\end{align}
On the other hand, we require that
the fields should be in the vacuum state \eqref{eq:VEV2} at the spatial infinity.
Therefore, we impose the conditions
\begin{align}
 u \to 0  ,~  f \to \eta  , ~ \alpha \to 0 \quad \mbox{as}\quad r\to \infty .
\label{eq:BC_infty}
\end{align}

\section{Numerical Solutions}

To obtain numerical solutions to the coupled ordinary differential equations \eqref{eq:eq_of_motion},
we use the relaxation method.
In numerics, we set $\eta=1$, and dimensional quantities are scaled as $r \to \eta r$,
$f\to \eta^{-1}f$, $u\to \eta^{-1}u$, $\alpha \to \eta^{-1}\alpha$,
and $\Omega \to \eta^{-1}\Omega$,  respectively.
We set $\lambda=1$, $e=1$, and $\mu=1.4$.

\begin{figure}[!ht]
\centering
\includegraphics[width=8cm]{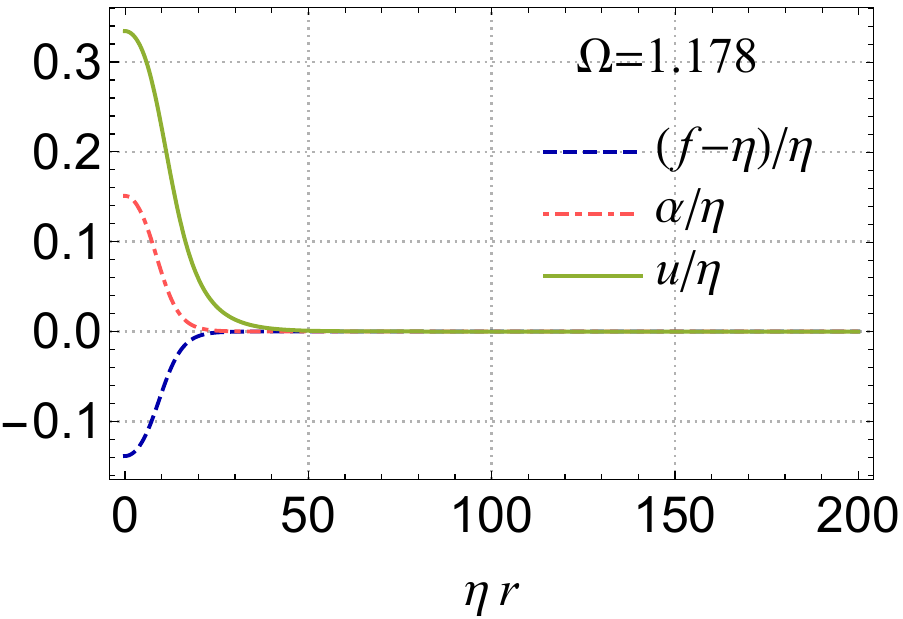}~~
\includegraphics[width=8cm]{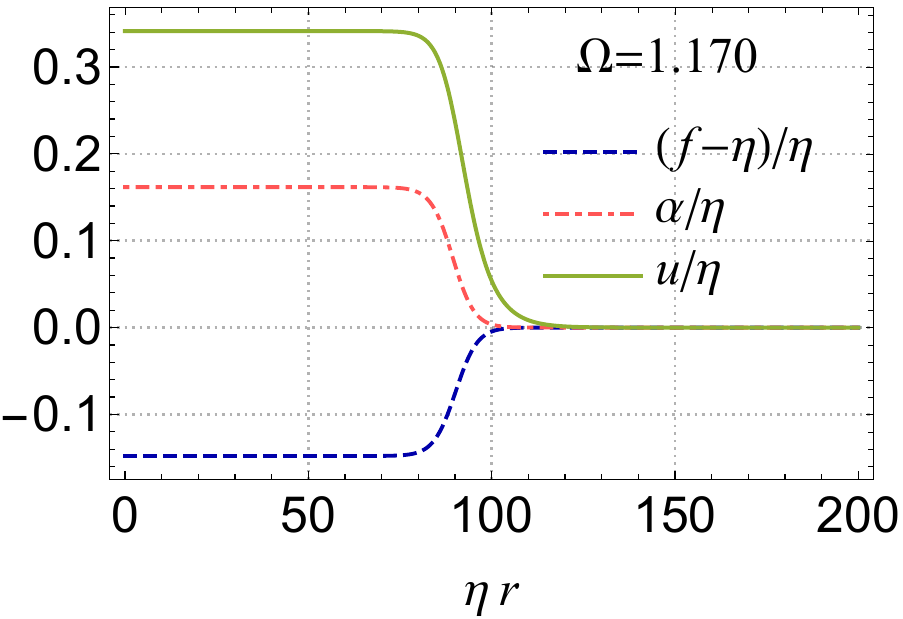}
\begin{minipage}[b]{0.9\hsize}
\caption{
Numerical solutions $u(r)$, $f(r)$, and $\alpha(r)$ are drawn for $\Omega=1.178$ (left panel),
and for $\Omega=1.170$ (right panel).
\label{fig:configurations}
}
\end{minipage}
\end{figure}

In Fig.\ref{fig:configurations}, we show numerical solutions
$u(r)$, $f(r)$, and $\alpha(r)$ for $\Omega=1.178$ and  $\Omega=1.170$, as examples.
In the both cases of $\Omega$, the functions are non-vanishing in a finite region,
and at large distances, \eqref{eq:BC_infty} is achieved.
Therefore, these solutions represent solitons.
In the case of $\Omega=1.178$, the functions are Gaussian function like,
while in the case of $\Omega=1.170$, the functions are step function like.
The soliton in the latter case represents a homogeneous ball,
all of the functions take constant values, $u(r)=u_0$, $f(r)=f_0$, and $\alpha(r)=\alpha_0$,
within a radius of the ball, and at the surface, $r=r_s$, the functions decay quickly.
This type of solutions are discussed by the thin wall approximation
in the literatures
\cite{Friedberg_Lee_Sirlin, Coleman, Lee_Stein-Schabes_Watkins_Widrow, Shi_Li, Gulamov_etal}.

\begin{figure}[!ht]
\centering
\includegraphics[width=8cm]{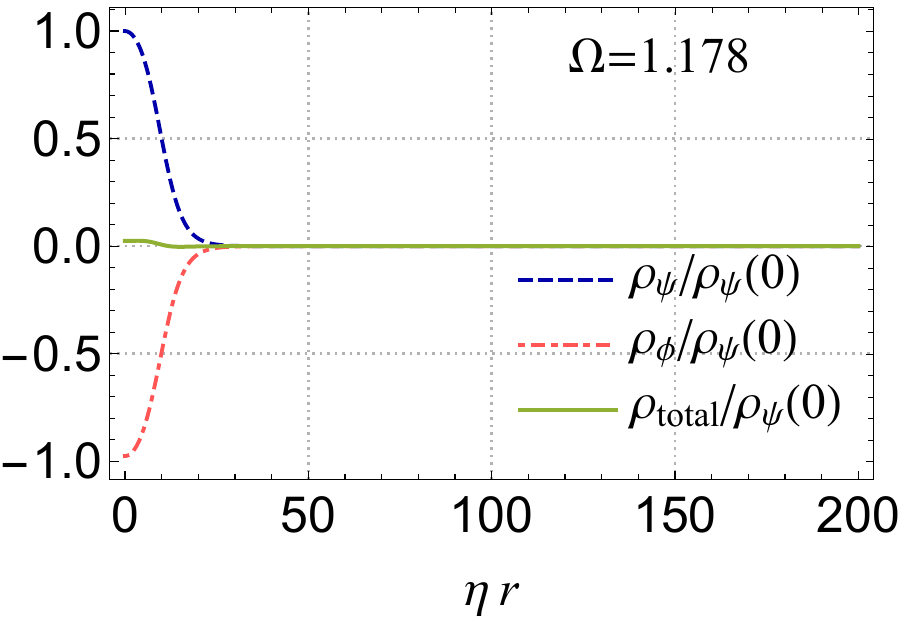}~~
\includegraphics[width=8cm]{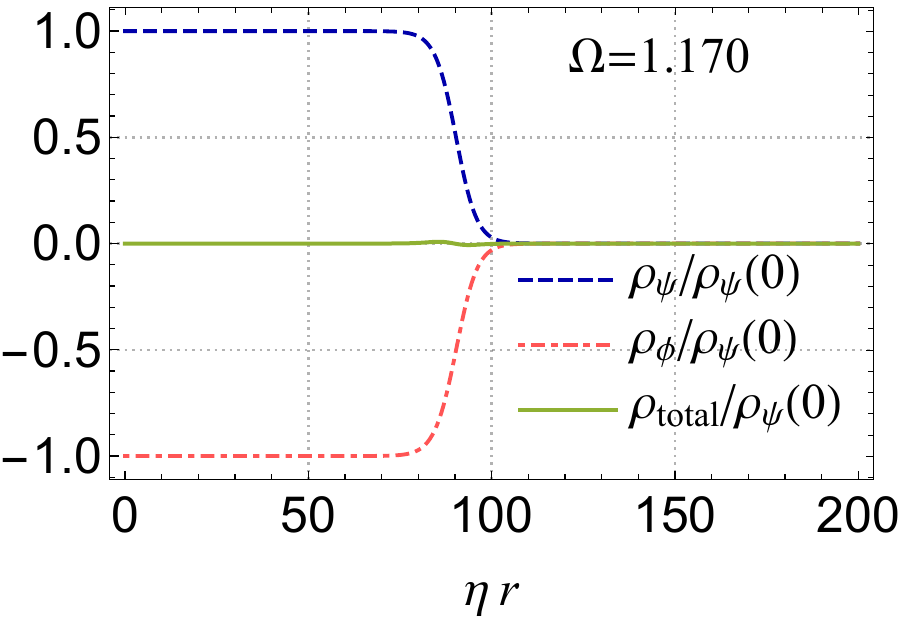}
\begin{minipage}[b]{0.9\hsize}
\caption{
The charge densities,  $\rho_{\psi}$, $\rho_{\phi}$, and $\rho_\text{total}:=\rho_\psi+\rho_\phi$
normalized by the central value of $\rho_{\psi}$
are shown as functions of $r$ for $\Omega=1.178$ (left panel), and for $\Omega=1.170$ (right panel).
\label{fig:charge mu0p35}
}
\end{minipage}
\end{figure}

We show the charge densities $\rho_{\psi}$, $\rho_{\phi}$
in Fig.\ref{fig:charge mu0p35} as functions of $r$.
In the both cases of $\Omega$, we find that $\rho_{\psi}$ is canceled out by
$\rho_{\phi}$, and the total charge density $\rho_\text{total}:=\rho_{\psi}+\rho_{\phi}$
almost vanishes everywhere, that is, the charge of the field $\psi$
is perfectly screened \cite{Ishihara:2018eah}.

\begin{figure}[!ht]
\centering
\includegraphics[width=8cm]{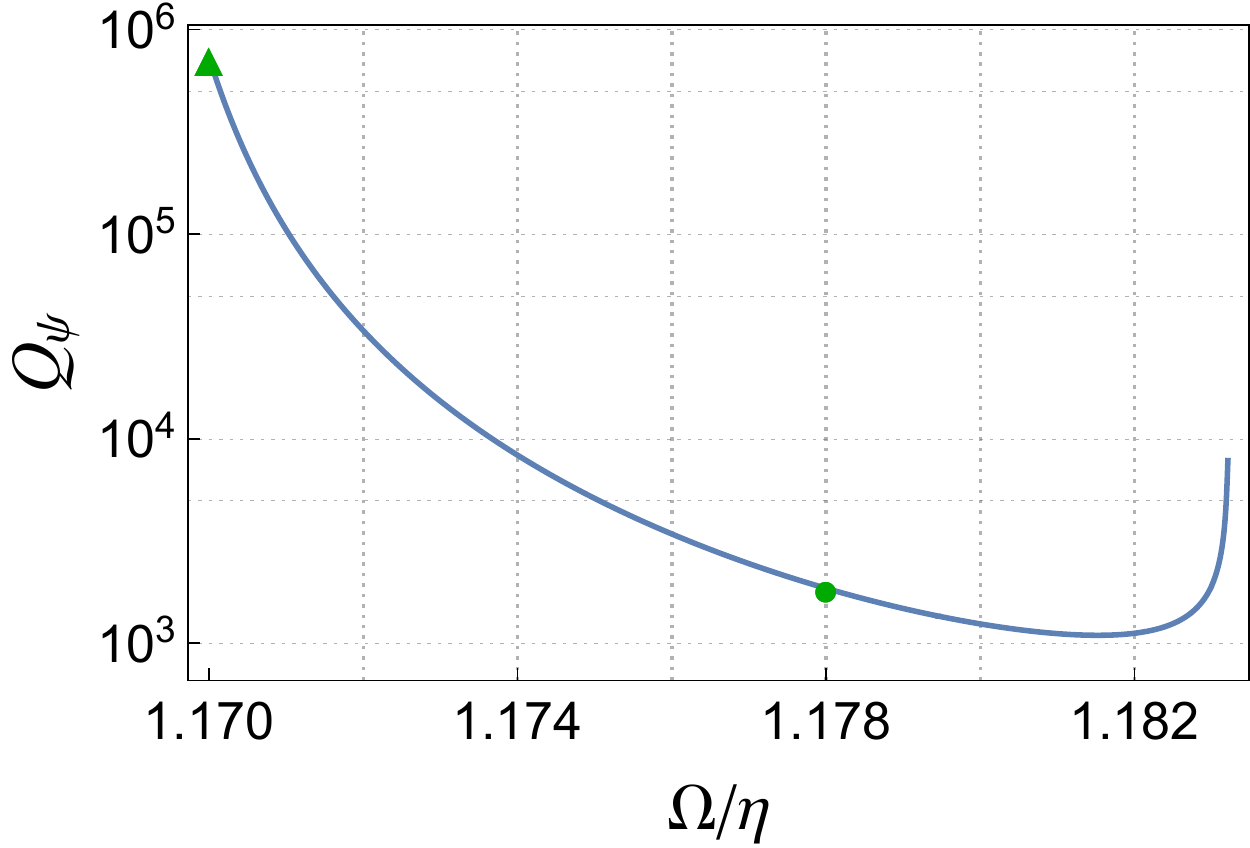}
\begin{minipage}[b]{0.9\hsize}
\caption{
The total charge of $\psi$, $Q_\psi$, is plotted as a function of $\Omega$.
$Q_\psi$ diverges at $\Omega=\Omega_\text{min}$ and $\Omega=\Omega_\text{max}$.
The circle in the figure, $\Omega=1.178$, corresponds to the left panel in Fig.\ref{fig:configurations},
while the triangle, $\Omega=1.170$, corresponds to the right panel.
\label{fig:Q_psi}
}
\end{minipage}
\end{figure}

The total charge of the scalar field $\psi$, $Q_\psi(=-Q_\phi)$, depends on $\Omega$
as shown in Fig.\ref{fig:Q_psi}.
The solution exists for $\Omega$ in the range $\Omega_\text{min} < \Omega <\Omega_\text{max}$,
where  $\Omega_\text{max}$ and $\Omega_\text{min}$ are given later.
At $\Omega=\Omega_\text{min/max}$, $Q_\psi$ diverges, respectively.

First, we determine $\Omega_\text{max}$.
Since $u$, $f-\eta$ and $\alpha$ are small at a large distance, solving the linearized
equations of \eqref{eq:eq_of_motion}, we have
\begin{equation}
\begin{split}
  	u(r)\propto  \frac{1}{r}\exp \left(-\sqrt{m_{\psi}^2-\Omega^2}~~r\right).
\end{split}
\label{eq:asymptotic_inf}
\end{equation}
If we require the solutions are localized in a finite region,
the parameter $\Omega$ should satisfies
\begin{align}
  \Omega^2 < \Omega_\text{max}^2 =m_{\psi}^2.
 \label{eq:Omega_max}
\end{align}

\begin{figure}[!ht]
\centering
\includegraphics[width=10cm]{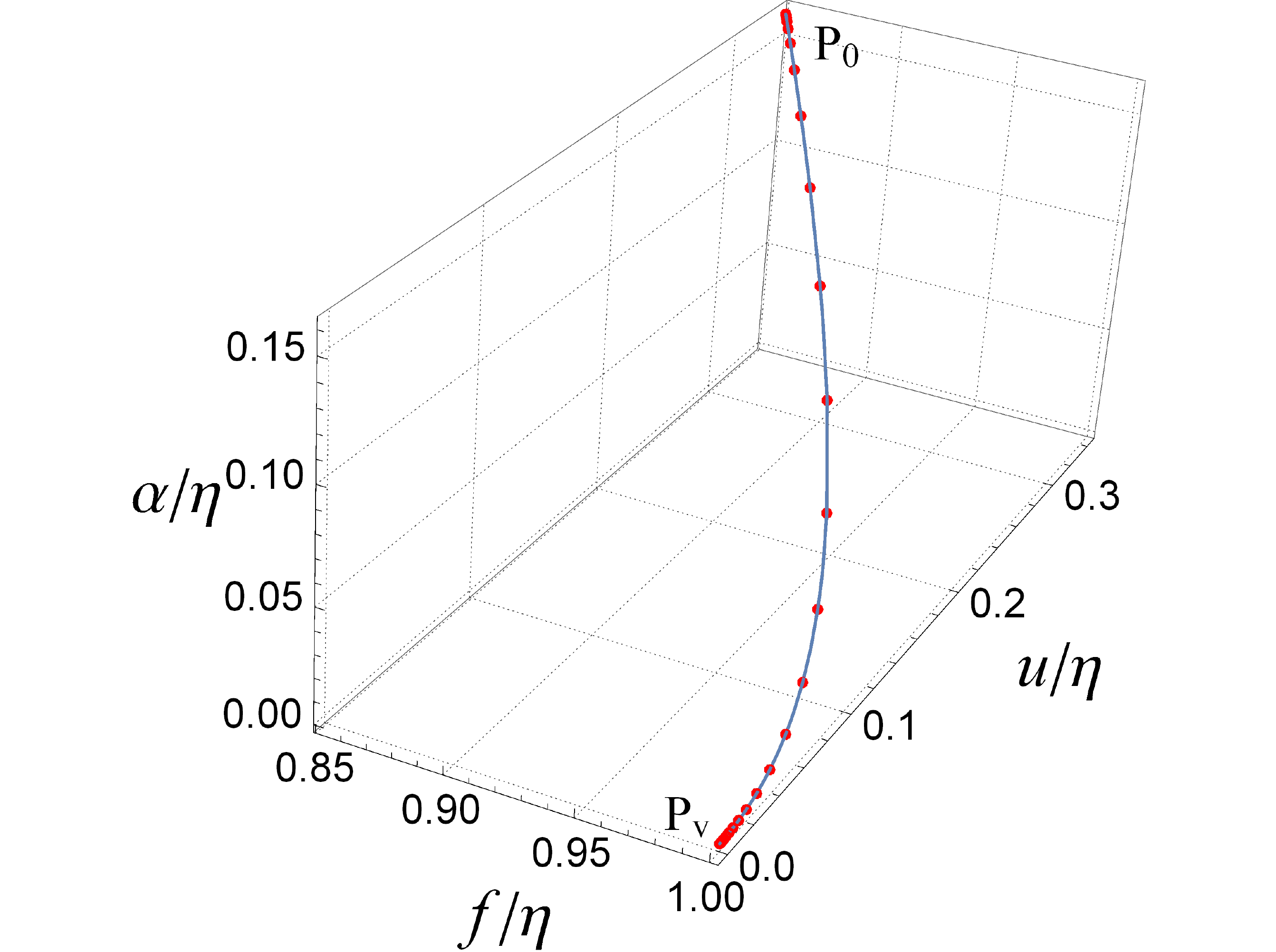}
\begin{minipage}[b]{0.9\hsize}
\caption{
Trajectory of the numerical solution for $\Omega=1.170$ in the $(u, f, \alpha)$
space. It starts from ${\rm P_0}$ and ends at ${\rm P_v}$. Dots on the trajectory
denote laps of $r$.
}
\label{fig:trajectory}
\end{minipage}
\end{figure}

Next, we determine $\Omega_\text{min}$.
If $\Omega$ takes a value near $\Omega_\text{min}$,
we have a homogeneous ball solution as shown in the right panel of Fig.\ref{fig:configurations}
as an example.
As $\Omega$ approaches to $\Omega_\text{min}$, $Q_\psi$ increases as same as
the radius of the ball increases very much. The homogeneous ball solution with a
large radius is described by a bounce solution of \eqref{eq:eq_of_motion},
a point in the three-dimensional space $(u, f, \alpha)$ sits on a stationary point of
the potential $U_\text{eff}$, say ${\rm P_0}$, long time,
and it moves to another stationary point, ${\rm P_v}$, that is the true vacuum in a short period,
and stays there finally. A trajectory of the solution in $(u, f, \alpha)$ space for $\Omega=1.170$
is shown in Fig.\ref{fig:trajectory}.
The stationary point ${\rm P_v}$ exists at $(u, f, \alpha)=(0, \eta, 0)$,
and ${\rm P_0}$ does at $(u, f, \alpha)=(u_0, f_0, \alpha_0)$, where
\begin{equation}
\begin{split}
	&\alpha_0 =\frac{1}{e(4\mu-\lambda)}
		\left((\mu-\lambda)\Omega+\sqrt{\mu(2\lambda+\mu)\Omega^2
		-\mu\lambda(4\mu-\lambda)\eta^2}\right),
\\
	&f_0 = \frac{1}{\sqrt{\mu}}(\Omega -e\alpha_0),
\quad
 	u_0 = \frac{1}{\sqrt{\mu}}\sqrt{e\alpha_0(\Omega -e\alpha_0)}.
\end{split}
  \label{eq:central_values}
\end{equation}
Here, $0 < e \alpha_0 <\Omega$ should hold for real value of $u_0$.
This condition with \eqref{eq:Omega_max} requires $\lambda < \mu$.
%
%
For the homogeneous ball solution with a large radius, the damping terms,
first derivative terms which are in proportion to $1/r$, in \eqref{eq:eq_of_motion}
are negligible around $r=r_s$.
In this case,
\begin{align}
	E_\text{eff}:=\biggl(\frac{df}{dr}\biggr)^2
		+ \biggl(\frac{du}{dr}\biggr)^2 -\frac12 \biggl(\frac{d\alpha}{dr}\biggr)^2
		+ U_\text{eff}(u,f,\alpha)
\end{align}
is conserved during the \lq evolution\rq\ in $r$, and then the bounce solution appears
if the potential height at the two stationary points are same, i.e.,
\begin{align}
	U_\text{eff}({\rm P_v} )=U_\text{eff}({\rm P_0} ).
\label{eq:U_stationary}
\end{align}
From the numerical calculations, we see that this occurs for $\Omega_\text{min}$.
Solving \eqref{eq:U_stationary}, we have
\begin{align}
  \Omega_\text{min} =\sqrt{m_{\phi}(2m_{\psi}-m_{\phi})}.
\label{eq:Omega_min}
\end{align}
Therefore, the allowed range of $\Omega$, $\Omega_\text{min}<\Omega<\Omega_\text{max}$ is rewritten as
\begin{align}
	2m_{\psi}m_{\phi}-m_{\phi}^2 < \Omega^2 < m_\psi^2 ,
\end{align}
equivalently,
\begin{align}
	2\sqrt{\lambda\mu}-\lambda < (\Omega/\eta)^2 < \mu.
\end{align}
Then, the nontopological soliton solution exists
for the model parameters satisfying $\lambda < \mu$.

\begin{figure}[!ht]
\centering
\includegraphics[width=8cm]{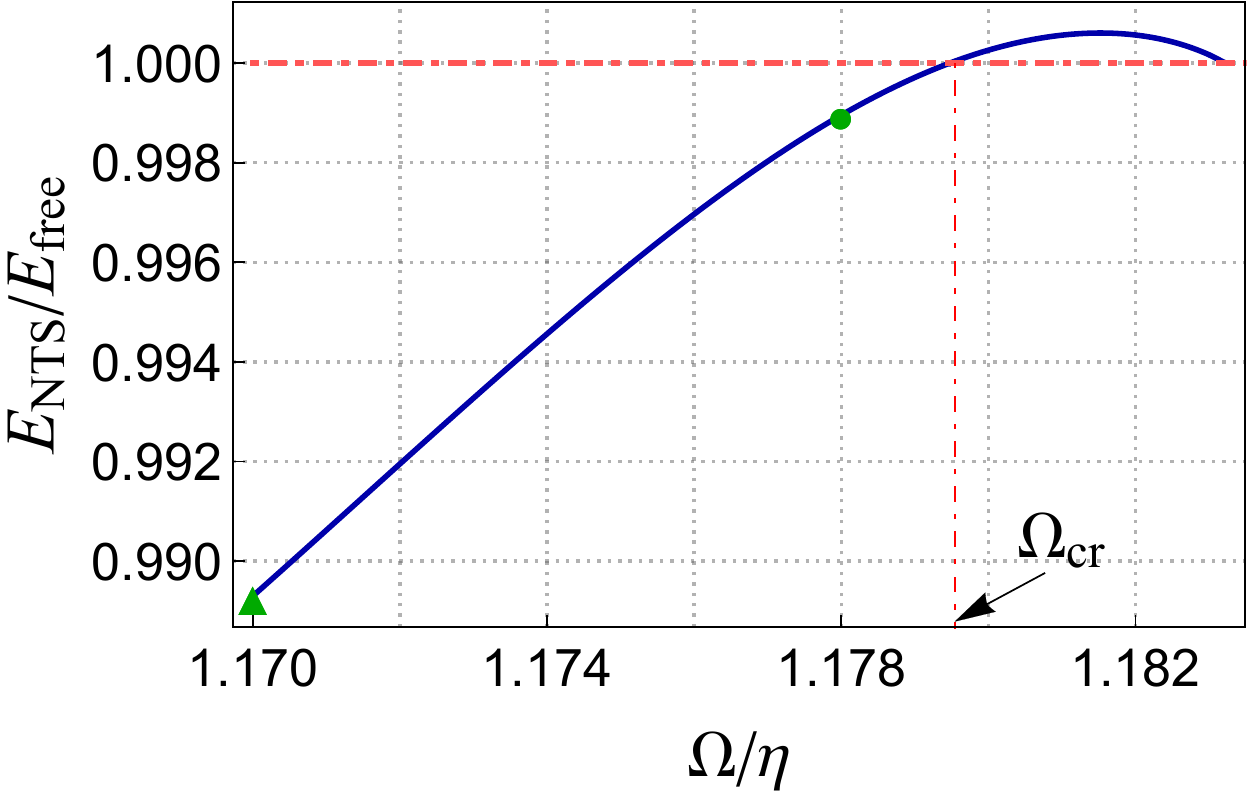}
\\
\begin{minipage}[b]{0.9\hsize}
\caption{The ratio $E_\text{NTS}/E_\text{free}$ as a function of $\Omega$.
For $\Omega_\text{min} < \Omega <\Omega_\text{cr}$, $E_\text{NTS}/E_\text{free}<1$.
The circle in the figure denotes $\Omega=1.178$,
while the triangle does $\Omega=1.170$.}
\label{fig:E_ratio_Omega}
\end{minipage}
\end{figure}

\begin{figure}[!ht]
\centering
\includegraphics[width=8cm]{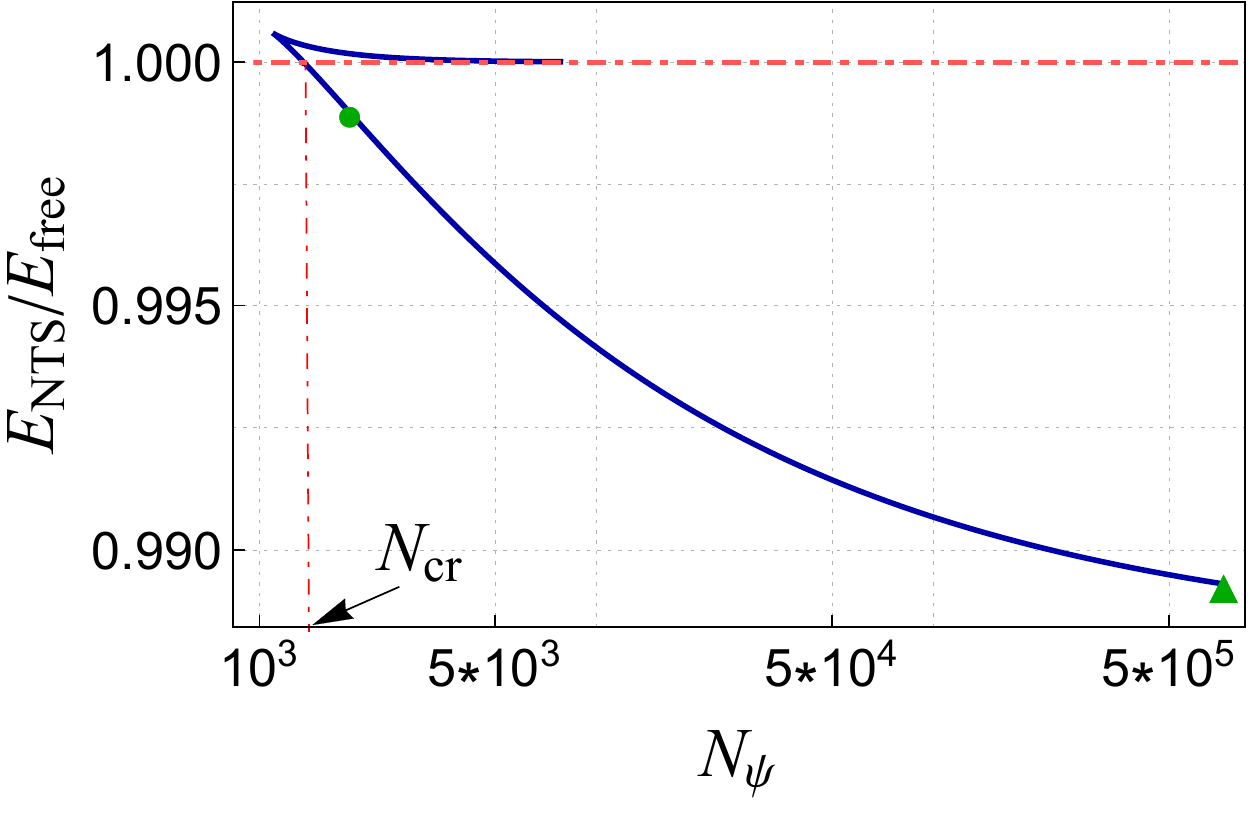}%
\\
\begin{minipage}[b]{0.9\hsize}
\caption{The ratio $E_\text{NTS}/E_\text{free}$ as a function of $N_\psi$.
The branch in the region $E_\text{NTS}/E_\text{free}<1$ corresponds to
$\Omega_\text{min} < \Omega <\Omega_\text{cr}$.
For $ N_\text{cr}<N_\psi$ on the branch, $E_\text{NTS}/E_\text{free}<1$.
The circle in the figure denotes $\Omega=1.178$,
while the triangle does $\Omega=1.170$. }
\label{fig:E_ratio_N}
\end{minipage}
\end{figure}

The nontopological soliton obtained in this paper can be regarded as a condensate of particles of
the scalar field $\psi$,
where the Higgs field plays the role of glue against repulsive force
by the $U(1)$ gauge field.
We compare energy of the soliton, $E_\text{NTS}$, given by \eqref{eq:energy2}
with mass energy of the free particles of $\psi$.
The numbers of the particles is defined by
\begin{align}
  	N_\psi &:=|Q_{\psi}|/e ,
  \label{eq:numbers_psi}
\end{align}
so that the free particles, as a whole, have the same amount of charge of the soliton.
Then, the mass energy of the free particles of $\psi$ is given by
$E_\text{free} = m_{\psi} N_{\psi}$.

In Fig.\ref{fig:E_ratio_Omega},
we plot the ratio $E_\text{NTS}/E_\text{free}$ as a function of $\Omega$.
We see that there exists $\Omega_\text{cr}$ such that
$E_\text{NTS}/E_\text{free}<1$ in the range
\begin{align}
	\Omega_\text{min}<\Omega<\Omega_\text{cr}.
\label{eq:Omega_cr}
\end{align}
The nontopological soliton in the range \eqref{eq:Omega_cr}
is preferable energetically, then the soliton does not decay into free particles.
In Fig.\ref{fig:E_ratio_N},
$E_\text{NTS}/E_\text{free}$ is plotted as a function of $N_\psi$.
We see that there exists a lower limit of numbers of condensed particles for
stable solitons, $N_\text{cr}$, but no upper limit.
A stable nontopological soliton with any large amount of mass possibly exists.
This is a significant difference from the case of gauged nontopological soliton
whose charge is not screened.

\section{Summary and Discussions~}

In this article, we have shown that nontopological solitons exist stably in the system
consisting of a complex scalar field coupled to a U(1) gauge field, and a complex Higgs scalar
field that causes spontaneously symmetry breaking.
The characteristic property of the solitons in this system is
the perfect charge screening \cite{Ishihara:2018eah},
namely, cancellation of the charge density of the complex scalar field by the counter charge cloud
of the other fields everywhere.
This is a desirable property for the nontopological solitons to be dark matter
\cite{Kusenko:2001vu, Kusenko:1997si, Kawasaki:2008zz}.
It is interesting that how much amount of the charge screened nontopological solitons produced during
the evolution of the early stage of the universe
\cite{Frieman:1988ut, Griest:1989bq, Kasuya:2000wx, Hiramatsu:2010dx}.

Owing to the perfect charge screening, infinitely heavy soliton is allowed in this system.
Then, two solitons would merge by collisions and form a larger soliton \cite{Kusenko:1997si},
and solitons with a astrophysical scale would appear finally.
It is important to investigate the gravitational effects of large solitons:
soliton stars, seeds of supermassive black holes,
and so on \cite{Friedberg:1986tp, Friedberg:1986tq, Lynn:1988rb, Mielke:2002bp}.

The system considered here would be embedded in more realistic field theories.
Generalization of the model is an interesting issue.
Furthermore, stability of the solutions should be clarified from various
points of view \cite{Cohen:1986ct, Kusenko:1997ad,
Multamaki:1999an, Paccetti:2001uh, Sakai:2007ft}.

\section*{Acknowledgements ~}
We are grateful to K.-i.Nakao, H.Itoyama, Y.Yasui, N.Maru, N.Sakai, and M.Minamitsuji
for valuable discussions.
H.I. was supported by JSPS KAKENHI Grant Number 16K05358.



\end{document}